# The Emergence of Spacetime: Transactions and Causal Sets

R. E. Kastner

7 November 2014

ABSTRACT. This paper discusses how the transactional interpretation of quantum mechanics can provide for a natural account of the emergence of spacetime events from a quantum substratum. In this account, spacetime is not a substantive manifold that becomes occupied with events; rather, spacetime itself exists only in virtue of specific actualized events. This implies that spacetime is discrete rather than continuous, and that properties attributed to spacetime based on the notion of a continuum are idealizations that do not apply to the real physical world. It is further noted that the transactional picture of the emergence of spacetime can provide the quantum dynamics that underlie the causal set approach as proposed by Sorkin and others.

1. Introduction and Background.

The transactional interpretation of quantum mechanics (TI) was first proposed by John G. Cramer (1986). Cramer showed how the interpretation gives rise to a physical basis for the Born Rule for probabilities of measurement outcomes. TI was originally inspired by the Wheeler-Feynman (WF) time-symmetric, 'direct action' theory of classical electrodynamics (Wheeler and Feynman 1945, 1949). The WF theory proposed that radiation is a time-symmetric process, in which a charge emits a field in the form of half-retarded, half-advanced solutions to the wave equation, and the response of absorbers combines with that primary field to create a radiative process that transfers energy from an emitter to an absorber. Davies later developed a quantum relativistic version of the WF Theory (Davies 1971, 1972). The present author has extended Cramer's TI into the relativistic domain based on the Davies theory (Kastner 2012). An additional element of this extension is to take quantum states and their interactions as describing pre-spacetime possibilities, rather than as process occurring in spacetime. This new version of TI is called 'Possibilist Transactional Interpretation' or PTI.

It should perhaps be noted that the direct action picture of fields has historically been somewhat neglected. This has been due not only to its counterintuitive time symmetric character, but also on the basis that the fields are not quantized, and therefore the direct action formalism is not convenient for practical computations. But it is also well known that quantum field theory is beset with serious mathematical consistency and conceptual problems; notably Haag's Theorem[1] (as well as the divergences requiring renormalization). It is therefore certainly possible that Nature's actual behavior is best described by the direct-action theory, even though it is not convenient for practical calculations.

The basic entities of TI are the 'offer wave' (OW), the retarded solution that corresponds to the usual quantum state $|\Psi>$ emitted by a source, and the 'confirmation wave' (CW), the advanced solution $<X|$. The CW is the response of absorber X to the component of the offer wave $|\Psi>$ projected onto the state $|X>$. As discussed in Kastner (2012), Chapter 3, the response of a set of absorbers (A,B,C…) to an offer wave $|\Psi>$ yields a physical referent for von Neumann's 'Projection Postulate,' which specifies that under measurement a pure state $|\Psi>$ is transformed into a mixed state, i.e.:

$$|\Psi><\Psi| \quad \rightarrow \quad \Sigma_i \, |<\Psi|X_i>|^2 \, |X_i><X_i| \quad , \tag{1}$$

where the weight of each projection operator corresponding to outcome $X_i$ is just the Born Rule. This process is illustrated in Figure 1.

---

[1] An informative discussion of Haag's Theorem and the challenge it poses for QFT is found in Earman and Fraser (2006).

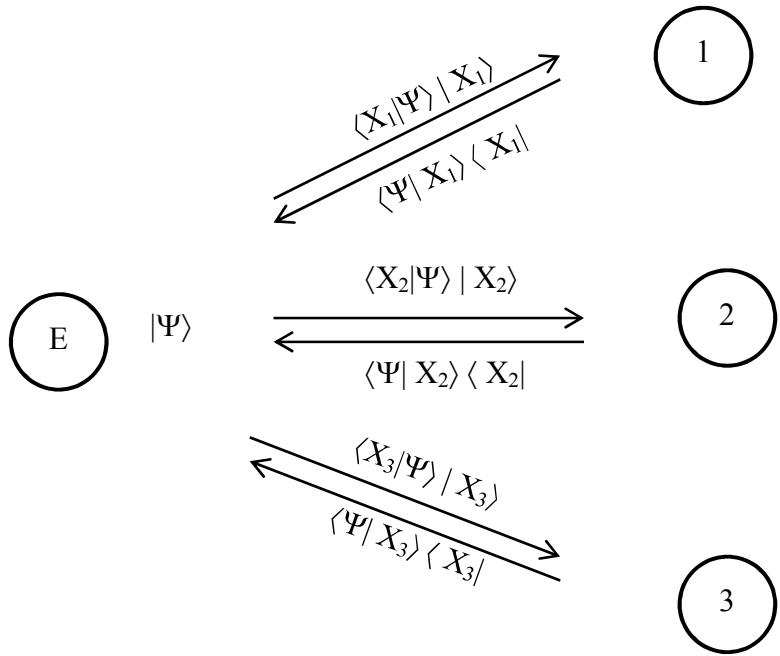

**Figure 1. An offer wave $|\Psi\rangle$ can be resolved into to various components corresponding to the properties of absorbers 1,2,3,… The product of a particular OW component $\langle X_i|\Psi\rangle |X_i\rangle$ with its corresponding CW component $\langle\Psi|X_i\rangle\langle X_i|$ reflects the Born Rule which tells us that the probability of the result corresponding to the projection operator $|X_i\rangle\langle X_i|$ is equal to $\langle X_i|\Psi\rangle\langle\Psi|X_i\rangle = |\langle X_i|\Psi\rangle|^2$.**

PTI adopts this basic formulation and extends the transactional picture into the relativistic domain by identifying the coupling amplitudes between fields as the basic amplitude for an offer (or confirmation) to be generated (see Kastner 2012, Chapter 6 and Kastner 2014). In addition, PTI proposes a growing universe picture, in which actualized transactions are the processes by which spacetime events are created from a substratum of quantum possibilities. The latter are taken as the entities described by quantum states (and their advanced confirmations); and, at a subtler relativistic level, the virtual quanta.

In PTI, what we call 'spacetime' is no more and no less that the causally connected collection of emission and absorption events corresponding to actualized transactions. Each actualized transaction defines a timelike (or null) spacetime interval whose endpoints are the emission and absorption. The emission is always in the past with respect to the absorption; the relationship between these two events corresponds directly to the 'link' in the causal set picture (described further below).

If a transaction involves a photon, the interval is null; if it involves a quantum with finite rest mass, the interval is timelike. The intervals have a causal relationship in that an absorption event A can, and generally does, serve as the site of a new emission event B. Thus the set of intervals created by actualized transactions establish a causal network with a partial order, much like the causal set structure proposed by Sorkin (2003). (The term 'causal set' is often abbreviated as 'causet'.) We address the specifics of the causet picture in the next section, but at this point, it is interesting to note the similar antisubstantival picture in Sorkin's presentation:

> A basic tenet of causet theory is that spacetime does not exist at the most fundamental level, that it is an "emergent" concept which is relevant only to the extent that some manifold-with-Lorentzian-metric $M$ furnishes a good approximation to the physical causet $C$.
> (Sorkin 2003, p. 9, preprint version)

An important feature of PTI is its relativistic extension of the basic transactional picture. This extension gives an account of the generation of offer waves, as an inherently stochastic process, from the direct action theory of quantum fields (cf. Davies 1971, 1972). This author has proposed, independently of the Sorkin's work on the causet picture, that this process is inherently Poissonian (i.e. based on decay rates). The basic idea is that offers and confirmations are spontaneously elevated forms of virtual quanta, where the probability of elevation is given by the decay rate for the process in question. In the direct action picture of PTI, an excited atom decays because one of the virtual photon exchanges ongoing between the excited electron and an external absorber (e.g. electron in a ground state atom) is spontaneously transformed into a photon offer wave that generates a confirming response. The probability for this occurrence is the product of the QED coupling constant $\alpha$ and the associated transition probability (see Kastner 2014). In quantum field theory terms, the offer wave corresponds to a 'free photon' or excited state of the field, instantiating a Fock space state.[2]

---

[2] However, the direct action theory itself does not assume an independently existing set of field oscillators, which allows it to escape the problems associated with Haag's theorem; this issue will be explored in a separate work.

When this process occurs, a set of incipient transactions is generally set up, as more than one absorber is generally available to any emitted photon offer wave. Each incipient transaction represents a choice of momentum direction for the emitted photon, which is emitted as a spherical (isotropic) wave. The Born Rule gives the probability that any particular incipient transaction will be actualized, but with certainty one of them will be actualized. Thus, when decay occurs, a new spacetime interval will be created. This corresponds to a new causally related pair of spacetime events; the emission event is the ancestor, and absorption event is the descendant. Thus, the Poissonian decay rates directly give rise to the sprinkling of new spacetime events of the kind envisioned in growing spacetime causal sets. We now turn to that formulation.

## 2. Causal Sets

The motivation for the causal set program as an approach to the vexed problem of quantum gravity is described by Sorkin as follows:

> The causal set idea is, in essence, nothing more than an attempt to combine the twin ideas of discreteness and order to produce a structure on which a theory of quantum gravity can be based. That such a step was almost inevitable is indicated by the fact that very similar formulations were put forward independently in [G. 't Hooft (1979), J. Myrheim (1978) and L. Bombelli et al (1987)], after having been adumbrated in [D. Finkelstein (1969)]. The insight underlying these proposals is that, in passing from the continuous to the discrete, one actually gains certain information, because "volume" can now be assessed (as Riemann said) by counting; and with both order and volume information present, we have enough to recover geometry. (Sorkin 2003, p. 5)

A causal set (causet) C is a locally finite partially ordered set of elements, together with a binary relation $\prec$. It has the following properties:

(i) transitivity: $(\forall x, y, z \in C)(x \prec y \prec z \Rightarrow x \prec z)$
(ii) irreflexivity: $(\forall x \in C)(x \not\prec x)$
(iii) local finiteness: $(\forall x, z \in C)$ (cardinality $\{ y \in C \mid x \prec y \prec z \} < \infty$)

Properties (i) and (ii) together imply that the elements are acyclic, while (iii) specifies that the set is discrete rather than continuous. This naturally leads to a well-defined causal order of distinct events, which can be associated with the unidirectionality of temporal becoming. Again, in Sorkin's terms:

> the relationship x ≺ y … is variously described by saying that x *precedes* y, that x is an *ancestor* of y, that y is a descendant of x, or that x lies to the *past of* y (or y to the *future* of x). Similarly, if x is an *immediate* ancestor of y (meaning that there exists no intervening z such that x ≺ z ≺ y) then one says that x is a *parent* of y, or y a *child* of x, …or that x ≺ y is a *link*. (Sorkin 2003, p. 7)

Again, as alluded to above, an actualized transaction defines a 'parent/child' relationship or link. Elements connected by such links are said to be *comparable*, or members of a *chain*.

Sorkin discusses how to create a causal set structure as a 'coarse-graining' of a continuous spacetime manifold M. The fundamental volume element of M corresponds to a single causal set element of C, so the basic correspondence between a causet C and a continuous manifold M is that N=V (where N is the number of causet elements approximating the volume V). In this context, he further notes:

> Given a manifold M with Lorentzian metric $g_{ab}$ (which is, say, globally hyperbolic) we can obtain a causal set C(M) by selecting points of M and endowing them with the order induced from that of M (where in M, x ≺ y iff there is a future causal curve from x to y). In order to realize the equality N = V, the selected points must be distributed *with unit density* in M. One way to accomplish this (and conjecturally the *only* way!) is to generate the points of C(M) by a *Poisson process*. (Sorkin 2003, p. 9)

As noted in the previous section, it was independently argued (Kastner 2014) that transactions are generated via decays, either of atomic excited states (which generate photon offers and confirmations) or of unstable nuclei (which generate offers and confirmations of quanta with nonvanishing rest mass). Such decay processes are always Poissonian. We return to the comparison between causets and the possibilist transactional process in Section 4.

3. Timelike and spacelike relations in the causet

A timelike relationship (i.e. either of ancestry or descendancy) obtains between elements of the causet that are *comparable*; that is, they are members of a single chain. On the other hand, a spacelike relationship obtains among elements that are all mutually incomparable; such elements are said to constitute an *antichain*. These relations between elements of a causet can be represented in a Hasse diagram, an example is shown in Figure 2.

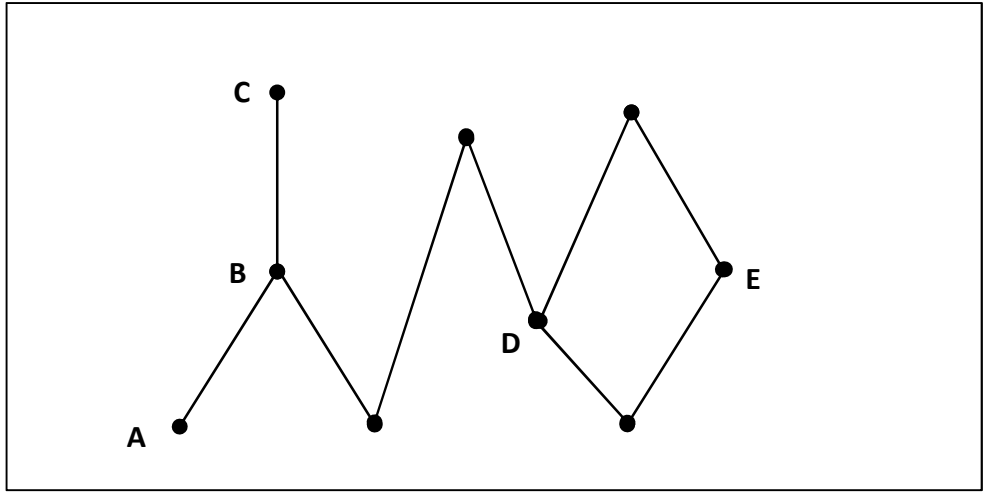

Figure 2. A simple example of a causet. Events are represented by dots and links by lines. The relation of descendance is indicated by the upward direction. Events A,B, and C are members of a chain, while events B, D and E are members of an antichain.

In the causet formulation, one cannot define spatial measure in terms of the structure 'orthogonal' to the chain; i.e., the antichain. The elements of an antichain by definition have no relationship to each other at all, and of course, there is no way to measure any aspect of a relationship where none exists. This rather strange feature is actually harmonious with the PTI account, in the following sense. In PTI, just as in relativity, only the spacetime interval has invariant physical content. On the other hand, temporal and spatial relationships are secondary, frame-dependent notions. These are only definable *with respect to a specific actualized transaction, as described in a particular frame*.

Since an actualized transaction is a necessary condition for definition of a spatial relationship between emitter and absorber, and an actualized transaction necessarily implies a temporal relationship (emission being the ancestor of the absorption), spatial displacement only obtains where there is also temporal displacement. That is, a temporal relationship must hold for any spatial relationship to be defined, even a frame-dependent one. Thus, space only exists when time exists; the concept of space has no physical meaning without a temporal relationship. However, a temporal displacement can be defined without a spatial displacement—the latter corresponds to a transaction viewed from the reference frame of a transferred quantum with finite rest mass.

The basic point is that we should not be surprised if it is difficult to define a purely 'spacelike' entity in the causet model. This should not be viewed as a weakness of the model but rather as a reflection of the fact that spatial relationships are supervenient both on temporal relationships and on frames of reference. Another way to put this is that no two events are ever truly 'simultaneous.' If they are not related by a chain (i.e., if they have no temporal relationship) then they cannot be regarded as having any spatial relationship either, including that usually implied by simultaneity.

4. Dynamics and growth of the causet

In the PTI picture, the growth of the causet is dictated by the underlying quantum dynamics. This of course presents a difficulty if one assumes that the time arguments in evolving quantum states $|\Psi(t)\rangle$ necessarily refer to spacelike hypersurfaces. The latter correspond to antichains in causet theory, and we just noted that one cannot define a spatial measure on these entities. However, the assumption that time indices refer to spacelike hypersurfaces is not in fact a necessary one. In what follows, we explore an alternative approach to the understanding of references to time in time-dependent quantum states.

The first point is that the Hamiltonian governing such evolving states is a "stand-in" for the net effect of scattering processes, which are mediated by quantum fields at the relativistic level. The Hamiltonian formulation is not relativistically covariant, since it singles out a preferred time coordinate. Thus we should not be surprised if the usual nonrelativistic time-dependent quantum state $|\Psi(t)\rangle$ seems incompatible with the relativistic causal set spacetime model; it is already incompatible with ordinary relativistic spacetime! Henson further comments that "… Even the Feynman path integral crucially refers to states on spacelike hypersurfaces." (Henson 2006, p. 9). However, the path integral formulation of first-quantized, nonrelativistic quantum mechanics also cannot be expected to be perfectly harmonious with a fully relativistic model of spacetime.

The way to address this issue is to view the time index in $|\Psi(t)\rangle$ as playing a conditional and relational role rather than an absolute one. Specifically, given the relevant potentials, $|\Psi(t)\rangle$ describes the nature of the offer wave that *would* be responded to by an absorber, such that the absorption event in a transaction actualized between the emitter and that absorber would be recorded at time *t* defined by reference to a clock in the absorber's frame (i.e. the absorber's proper time).

To understand this conditional nature of the time index, recall that the dynamical evolution in a time-dependent state such as $|\Psi(t)\rangle$ is given by the relevant Hamiltonian: $|\Psi(t)\rangle = \exp(-iHt/\hbar) |\Psi(0)\rangle$, where $|\Psi(0)\rangle$ is the emitted offer wave. As noted above, the Hamiltonian *H* in the time evolution operator, $U = \exp(-iHt/\hbar)$, describes the overall effect of relativistic scattering processes. Suppose it is projected that the offer wave (or 'quantum system' in the usual parlance) will reach a given macroscopic absorber when the laboratory clock reads $t = t_a$. The value of the time evolution operator at $t_a$ is a measure of the interactions of the applicable forces via scattering with the offer, and thus their net effects on the offer, with respect to that proper time interval. While such interactions are often assumed to be taking place in spacetime, that is not a necessary assumption.[3] It is rejected in PTI, which takes such processes as pre-spatiotemporal and sub-empirical. Indeed these processes are what underlie and give rise to the spacetime manifold which is the causet itself. Thus, only the locally measured proper times of emission and absorption can provide a physically relevant temporal measure of the evolution undergone by the quantum system between its emission and absorption.

How does this work? Consider again the Hasse diagram of Figure 2, which illustrates a particular stage of growth of the causet. We also have to consider the causet as being embedded in a quantum substratum of interacting emitters and absorbers (e.g. excited and unexcited atoms); this substratum is represented in Figure 3 by a patterned background. (Some of these atoms have very high probabilities of emitting to other atoms, and vice versa; such groups of mutually emitting and absorbing atoms comprise macroscopic objects.) A later stage of growth can be represented by the addition of a new additional event F, which arises from the actualization of a transaction between C (as emitter) and F (as absorber):

---

[3] It has been noted by Beretstetskii et al (1971, p. 3) and Auyang (1995, p. 48) that processes mediated by quantum fields are not appropriately viewed as spacetime processes. Specifically, Auyang notes that spacetime indices refer to points on the field, not spacetime points.

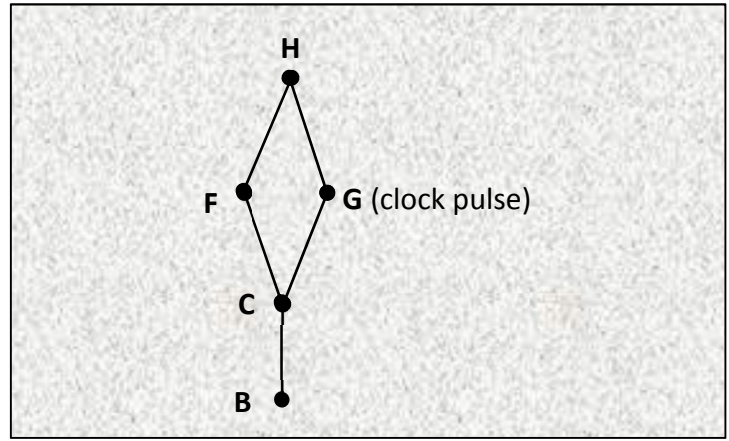

**Figure 3. A new event F is added to the causet. Its temporal relationship to the earlier event C can be inferred by reference to a clock pulse, shown as another new event G. These events must be causally connected at a later event H in order to infer the time interval between C and F.**

At the microscopic level, an object/event actualized as an absorber in one transaction, such as an atom labeled C in Figure 2, becomes re-actualized as an emitter in a succeeding transaction (as in Figure 3, where C emits to the absorber at F). The emission occurs as it decays from its previous excited state and emits a photon offer wave to the next absorber (i.e. another atom) actualized at F. Note again that this is a Poissonian process, which fulfills the requirement that event 'sprinkling' into the causet must be Poissonian to preserve relativistic covariance.

The time interval between events C and F can only be defined relative to a clock—i.e., relative to some pre-established periodic process.[4] This is indicated in Figure 3 by the chain segment from C to G, which counts one unit of time as measured by a relevant clock. If an identical transaction (i.e. conveying the same amount of energy) then takes place between F and a later event H, which serves also as a direct descendant of G, then we can infer that the time interval between C and F was one unit. This is not strictly possible at the microscopic level, since an absorber can only participate in one transaction at any instant. Thus the definition of a time interval at the microscopic level can only be approximate.

---

[4] An example is an atomic clock, which allows one to relate an atomic transition frequency to a unit of time by counting oscillations (as in those of the microwave oscillator driving a Cesium clock in resonance with the principal transition frequency). Such oscillations would constitute a causally connected set of transacted events -- a 'chain' in the causet with well-defined time intervals. (See Kastner 2012, Chapters 3 and 6, for details on how the transactional picture enables definition of the macroscopic realm, which would include objects such as a microwave oscillator.)

Nevertheless, in order to establish an empirical spacetime structure at the macroscopic level, it is not really required that the *same* atom absorb and then re-emit. It is sufficient that a macroscopic object absorbs and then re-emits, in which case the absorption and emission may be carried out by different atoms or molecules comprising the macroscopic object. As noted above, collections of atoms with high probabilities of continually emitting and absorbing to one another comprise macroscopic objects. (A simple example of this sort of absorption and re-emission process is a small macroscopic sample of gas whose molecules are undergoing continual thermal interactions; the latter are transactions.)

Figure 4 is a 'bare bones' model of a macroscopic absorber $F$ with a laboratory clock $G$ attached to it and causally connected to the macroscopic emitter $C$ as well:

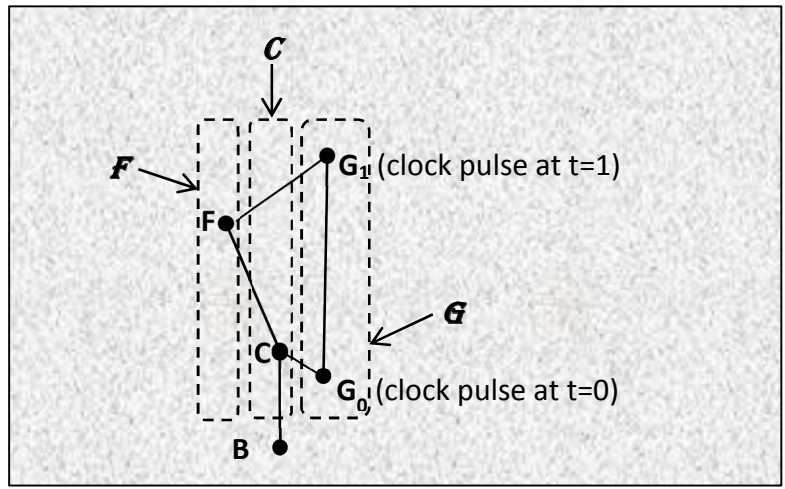

**Figure 4.** A macroscopic emitter $C$, macroscopic absorber $F$, and laboratory clock $G$ (world tubes indicated by dashed rectangles) are all causally connected via ongoing transactions with the laboratory equipment (only those between emitter/absorber, emitter/clock and absorber/clock are shown explicitly). The clock measures the proper times of the emission at C and the absorption at F. Note that there is an inherent limit to the accuracy of the measurement, since the absorptions are never strictly simultaneous. (Diagonal connecting lines indicate photon transactions at a speed of *c*.)

Thus, the temporal reference which appears as a challenge in developing the causet picture is actually an asset in the PTI model: we do not need to refer to a spacelike structure in the causet in order to apply quantum theory to the growth of the causet. Rather, we can understand time-dependent quantum theory as referring to an evolving entity (the changing offer wave) in the quantum substratum, which is not contained in the causet itself. Entities in the quantum substratum can undergo change without necessary reference to time, which applies only at the actualized spacetime level. The relevant time interval is then defined locally, with respect to the actualizing absorber and its interactions with other components (such as clocks). Thus, it is only through an actualized transaction that the evolving offer wave gains a well-defined temporal reference. An absolute time reference is inappropriate for the quantum object, since (1) the quantum object is a pre-spacetime (pre-causet) entity, and in any case (2) that would inevitably involve a hyperplane of simultaneity that cannot be reconciled with relativistic covariance.

5. Conclusion

The possibilist transactional picture can be viewed as a physical basis for the emergence of the partially ordered set of events in the causal set formalism. This formalism is currently being explored as a means to constructing a satisfactory theory of quantum gravity, and it has much promise in that regard. However, even apart from general relativistic considerations, the formalism breaks new ground in showing that, contrary to a well-entrenched belief, a block world ontology is not required for consistency with relativity. The causal set structure is a 'growing universe' ontology which nevertheless preserves the relativistic prohibition on a preferred frame.

Likewise, the transactional ontology proposed here is a variation on the 'growing universe' picture. The account is consistent with relativity theory in that the set of events is amenable to a covariant description: no preferred frame is required. This is because the transactional process is inherently Poissonian, and therefore preserves the relativistic covariance of the causal set model.